\documentclass{article}
\usepackage[margin=1in]{geometry}
\usepackage{graphicx}
\usepackage{bbm}
\usepackage{amsmath,amssymb,amsfonts}
\usepackage{algorithm}
\usepackage{algorithmicx}
\usepackage[noend]{algpseudocode}
\usepackage{textcomp}
\usepackage{xcolor}
\usepackage[caption=false, font=footnotesize]{subfig}
\usepackage{mathtools}
\usepackage{url}
\usepackage[T1]{fontenc} 
\usepackage{float}
\usepackage{arydshln}

\usepackage[acronym]{glossaries}

\newacronym{am}{AM}{affinity mixup}
\newacronym{amn}{AMN}{affinity mixup network}
\newacronym{ann}{ANN}{attention neural network}
\newacronym{crnn}{CRNN}{convolutional recurrent neural network}
\newacronym{fft}{FFT}{fast Fourier transform}
\newacronym{gnn}{GNN}{graph neural network}
\newacronym{gru}{GRU}{gated recurrent unit}
\newacronym{lms}{LMS}{log-Mel spectrogram}
\newacronym{mil}{MIL}{multiple instance learning}
\newacronym{nn}{NN}{neural network}
\newacronym{relu}{ReLU}{rectified linear unit}
\newacronym{sed}{SED}{sound event detection}
\newacronym{wssed}{WSSED}{weakly supervised sound event detection}

\usepackage{biblatex} 
\usepackage{setspace}
\usepackage{tikz}
\definecolor{spc}{HTML}{F5F5F5}
\definecolor{conv2d}{HTML}{B0E3E6}
\definecolor{downsample}{HTML}{FFE6CC}
\definecolor{am}{HTML}{E1D5E7}
\definecolor{rnn}{HTML}{B1DDF0}
\definecolor{linear}{HTML}{DAE8FC}
\definecolor{upsample}{HTML}{D5E8D4}
\definecolor{pooling}{HTML}{F8CECC}
\newcommand{\colorcircle}[1]{\tikz\draw[black,fill=#1] (0,0) circle (.9ex);}

\begin{document}

\begin{center}
\Large{\textbf{Affinity Mixup for Weakly Supervised Sound Event Detection}} \\
\vspace{10pt}
\large{Mohammad Rasool Izadi, Robert Stevenson, Laura N. Kloepper} \\
\vspace{10pt}
\end{center}

\section{Abstract}
The \acrfull{wssed} problem is the task of predicting the presence of sound events and their corresponding starting and ending points in a weakly labeled dataset. A weak dataset associates each training sample (a short recording) to one or more present sources. Networks that solely rely on convolutional and recurrent layers cannot directly relate multiple frames in a recording. Motivated by attention and graph neural networks, we introduce the concept of an \acrfull{am} to incorporate time-level similarities and make a connection between frames. This regularization technique mixes up features in different layers using an adaptive affinity matrix. Our proposed \acrfull{amn} improves over state-of-the-art techniques event-F1 scores by $8.2\%$. 

\textit{Keywords}: Sound event detection, weakly supervised learning, attention, graph

\section{Introduction} \label{sec:introduction}
\Acrfull{sed} aims to provide two types of information about an audio recording: the presence of known sources and their corresponding onset and offset times (the starting and ending points); The former task is usually referred to as tagging, while the latter is called localization. Some of the applications of \acrshort{sed} include context-based indexing and retrieval in multimedia databases, unobtrusive monitoring in healthcare, and audio-based surveillance \cite{lee2014acoustic}. Detected events can also be used as mid-level representation in other research areas, e.g. audio context recognition \cite{heittola2010audio}, and automatic tagging and segmentation \cite{wichern2010segmentation}. For a recording of multiple (possibly overlapping) events, the problem, known as polyphonic \acrshort{sed} \cite{mesaros2016metrics}, is studied in various scenarios, such as synthetic or loosely labeled audio. In supervised learning, where both clip-level and frame-level labels are available, a \acrshort{sed} system potentially performs better than a \acrfull{wssed}, where the dataset provides only clip-level labels \cite{dinkel2021towards}. Since frame-level labels are certainly expensive to gather, \acrshort{wssed} has gained an increasing attention recently and the overall performance gap is closing. 
A common framework for \acrshort{wssed} is \acrfull{mil} in which the dataset labels are bags of classes \cite{babenko2008multiple}. In such datasets, all of the frame-level samples of a clip are labeled with a class if the clip contains at least one frame of that class. To date, \acrfullpl{crnn} with outputs for tagging and localization achieve the best results for the \acrshort{wssed} problem. The tagging output finds the probability of each frame belonging to each class. While the localization is a pooled version of the tagging output. The pooling function aggregates the frame-level probabilities into a clip-level probability. The cost function is set to be the distance between the bag of labels and clip-level probabilities. 

There are two main issues with encoding/up-sampling \acrshortpl{crnn}. The encoding step is a parametric function that reduces the spectrogram size multiple times and the up-sampling function is just a pre-defined function which generates the final frame-level predictions from the lowest resolution. Previous work has shown that simply replacing the up-sampling function with parametric layers (e.g. transposed convolutions) does not help with the \acrshort{sed} performance \cite{dinkel2021towards}. Since the up-sampling function is independent of the input, the encoder determines the final accuracy. In addition to the resolution issue with the up-sampling, a \acrshort{crnn} is not capable of fully propagating temporal information. Convolutional layers capture the time domain relations through small $2$D kernels. The recurrent layer models the sequence graph using hidden states/cells. However, there’s no global scenario for directly connecting features in time. To remedy this issue, \acrfullpl{ann} suggest routing the temporal features using a similarity matrix. In a similar approach to the message passing with normalized adjacency in \acrfullpl{gnn}, \acrshort{ann} models the temporal information with a scaled dot-product. However, in the \acrshort{wssed} problem, the adjacency is not available and the dot-product only works with normalized inputs.

We introduce the concept of the \acrshort{amn} to provide a mechanism to mix up time-level features based on their similarities in both encoding and decoding; If two features are very much alike, they should be closer. To bring similar samples together, an affinity matrix is constructed based on the Euclidean distance. The \acrshort{am} is a regularization technique using an affinity matrix which represents the relationship between samples. Unlike previous methods, a decoder replaces the up-sampling step. 

Section \ref{sec:background} formally defines the \acrshort{wssed} problem and provides necessary background on \acrshort{ann} and \acrshort{gnn}. The last part of the section elaborates on time and frequency resolutions. Section \ref{sec:method} introduces the proposed method in detail. This section also provides the relationship between \acrshort{am} and other similar methods (\acrshort{ann} and \acrshort{gnn}). The experimental setup and results are provided in section \ref{sec:experiments} and the work is concluded in section \ref{sec:conclusion}.

\section{Background} \label{sec:background}
\subsection{Weakly Supervised Learning}
In the supervised learning, the target function is estimated using a set of one-to-one examples, e.g., $x_i \rightarrow y_i$. Weakly supervised learning is a type of \acrshort{mil} in which a set of samples are tagged with a set of labels, e.g. $\{x_i, \dots\} \rightarrow \{y_k, \dots\}$. Having many-to-many relationships between the examples in the training, the aim of \acrshort{wssed} is to estimate two different functions: clip-level and frame-level. For a clip $X = (x_1, \dots, x_t)$ in a $c$-class scenario, the clip-level function $p(X) \in [0, 1]^c$ predicts the probability of $X$ containing each class (tagging) and the frame-level function $q(X) \in [0, 1]^{t \times c}$ measures the chance of each frame belonging to each class (localization). Note that a clip may contain multiple classes and a frame may belong to multiple classes. Since the training and evaluation of the clip-level function follows the same criteria, the tagging step is much easier than estimating frame-level function and it can be learned without a direct supervision. An immediate solution to this problem is to construct the clip-level $p(X)$ by pooling the the frame-level $q(X)$. In general, a pooling function is nothing but a weighted sum over the frames, i.e. $p(X) = a(X) q(X)$, for $a(X) \in [0, 1]^t$ and $\sum_i a(X) = 1$. In addition to the max pooling, $p(X) = \max_i q(X)_i$, and soft-linear, $p(X) = \sum q(X)^2 / \sum_i q(X)$, one can predict the $a(X)$ in an extra step which resembles the self-attention mechanism \cite{vaswani2017attention}.

\subsection{Attention and Graph Neural Networks}
\Acrshortpl{ann} and \acrshortpl{gnn} layers can be used to transform a feature without changing its duration while considering time domain similarities; i.e. $X \in \mathbbm{R}^{t \times f} \rightarrow X' \in \mathbbm{R}^{t \times f'}$. \acrshortpl{ann} were first proposed for natural language processing \cite{vaswani2017attention}, where the words in a sentence are attended differently for machine translation. In \acrfullpl{ann}, each feature attends to other similar features and ignores irrelevant ones. Let $U \in R^{f \times f'}$ and $V \in R^{f \times f'}$ be the similarity and output parameters, respectively. An \acrshort{ann} layer controls how much each time domain sample is related or should be attended and defined as
\begin{equation} \label{eq:att}
X' = \text{softmax}(\frac{(XU) \hspace{0.5em} (XU)^\top}{\sqrt{f'}}) XV.
\end{equation}
The \acrfull{gnn} extends the \acrfull{nn} mapping to the graph structured data \cite{9378063}. The basic idea is to use related samples when the adjacency information is available. When having the adjacency matrix $A \in [0, 1]^{t \times t}$ and its associated degree matrix $D \in \mathbbm{N}^{t \times t}$, a \acrshort{gnn} layer can be written as 
\begin{equation} \label{eq:graph}
X' = \phi(\tilde{A}X)
\end{equation}
for some element-wise non-linear function $\phi$ and the normalized adjacency $\tilde{A} = (D+I)^{-1/2}(A+I)(D+I)^{-1/2}$ where $I$ denotes a $t \times t$ identity matrix.
\subsection{Time and Frequency Resolutions}
The \acrshort{sed} problem starts with a single-channel clip $x \in \mathbbm{R}^s$ and ends with both clip-level and frame-level outputs. Throughout this process, the signal resolution and the number of channels change repeatedly. Instead of directly inputting ambient sound pressures, the front-end features in SED are typically Fourier transform of short frames. The Fourier transform introduces the frequency dimension with the cost of losing the time resolution. It can be seen as a $1$D convolution followed by an exponential where the number of outputs is set by the number of \acrfull{fft} points. Let $X \in \mathbbm{R}^{t \times f}$ be the front-end features in the time-frequency domain for $t < s$. Since the time-frequency features are the initial inputs to the parametric model, $t$ is the highest time resolution, i.e. $q(X) \in [0, 1]^{t \times c}$. A \acrshort{nn} layer not only changes the number of input channels but also alters time and/or frequency resolutions. A series of encoding layers, each containing a down-sampler, reduces the time resolution into $t' < t$. Note that $l$-norm down-sampling helps with the duration robustness more than common operators like average or max \cite{dinkel2020duration}. Shrinking the features across the time axis helps with reducing the number of steps in the recurrent layers and disjoint predictions like $(\dots, 1, 0, 1, 0, \dots)$. After the encoding, one or more up-sampling steps take the resolution $t'$ back to $t$. To this end, a single linear up-sampler typically performs better than parametric functions like transposed convolutions \cite{dinkel2021towards}.

\section{Affinity Mixup} \label{sec:method}
\begin{figure}[H]
\begin{minipage}[b]{1.\linewidth}
  \centering
  \centerline{\includegraphics[width=244pt]{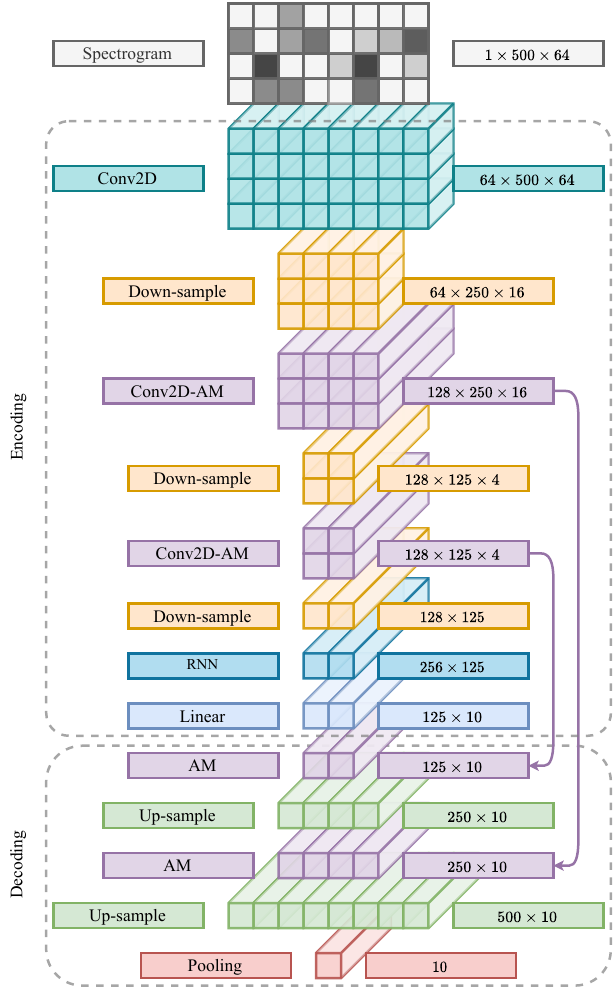}}
\end{minipage}
\vspace*{-15pt}
\caption{The \acrshort{amn} architecture is depicted for a spectrogram with the duration of $500$ frames and $64$ frequency bands. It has two principal components: encoding and decoding. A Conv2D block contains an initial batch normalization, then a same-size convolution, and finally, a leaky \acrshort{relu} (slope = $-0.1$) as the activation. A down-sample operation reduces time and/or frequency resolutions. The encoding ends with a linear layer projecting to the output space of $c=10$ classes. The model has two outputs: time-level (the second up-sample) and clip-level (pooled time-level). The training is done using only the clip-level output and the time-level is used for the evaluation.}
\label{fig:net}
\end{figure}

Figure \ref{fig:net} shows the proposed approach, a \acrshort{crnn} architecture with two \acrshort{am} layers. The spectrogram \colorcircle{spc} is the time-frequency representation of sound pressure measurements in terms of energy/power. To start from the time domain, one can replace the spectrogram with a Conv$1$D followed by an exponential where the number of outputs is set by the number of \acrshort{fft} points. The \acrshort{sed} is performed in two steps of encoding and decoding. The encoder part includes three parametric layers; A Conv$2$D \colorcircle{conv2d} and two Conv$2$D-AM \colorcircle{am}. Each layer is composed of a $2$D batch normalization, $2$D convolution, and non-linearities like (leaky) \acrfull{relu}. After each layer, time and/or frequency resolutions is reduced by pooling/down-sampling layers \colorcircle{downsample}.
The last part of the encoder is a bidirectional sequential layer \colorcircle{rnn} like \acrfull{gru} followed by a linear layer \colorcircle{linear} projecting features into the output space with $c=10$ classes. The time-level outputs, the probability of each class at each moment, are predicted using two up-sampling steps \colorcircle{upsample}. Finally, a pooling function \colorcircle{pooling} like the max or soft-linear predicts the clip-level outputs.

The \acrshort{am} is a two-step regularization that can be applied in every time resolution. In the first step, the resolution-dependent affinity matrix is calculated for an encoding feature. In the second step, it’s used for mixing up every other feature with the same time resolution. The affinity matrix can be constructed via a pair-wise distance matrix in several ways such as nearest neighbors, ball neighborhood, or kernel functions. However, since the affinity function requires to be diffrentiable, sorting-based methods like the nearest neighbors are not applicable to a \acrshort{nn} layer. Also, the ball neighborhood has an extra parameter that can be layer-dependent in general. So for simplicity, the affinity matrix is calculated just based on an exponential kernel function. Let $X \in \mathbbm{R}^{b \times t \times f}$ be an encoder feature where $b$ denotes the number of channels. Also, $t$ and $f$ show the time and frequency resolutions, respectively. In a $c$-class problem, the encoder is projected into $\tilde{X}=WX$ for $W \in \mathbbm{R}^{c \times b}$. Given a pairwise distance $\mathrm{d}: \mathbbm{R} \rightarrow \mathbbm{R}_{+}^{c \times t \times t}$, the affinity matrix is calculated by 
\begin{equation} \label{eq:am}
A = \text{softmax}(\frac{-\mathrm{d}(\tilde{X}, \tilde{X})^2}{\tau \sqrt{f}}).
\end{equation}
where $\tau$ is the only hyper-parameter. When having multiple features with the same time resolutions in a row, the affinity matrix is calculated using the first one.

Having the affinity matrix, it can be used to mix up any other feature with the same time resolution in the second step. In this work, for a $t$-resolution affinity matrix, there are two other features with the same time resolution; One in the encoder $X' \in \mathbbm{R}^{b' \times t \times f'}$ for $X' = \mathrm{f}(X)$ and one in the decoder $Z' \in [0, 1]^{t \times c}$ for $Z' = \mathrm{g}(Z)$. The function $\mathrm{f}$ in the encoder is the second or third conv2D block (including batch normalization and non-linearity). Respectively, the function $\mathrm{g}$ in the decoder is the first or second identity function before up-sampling (see Fig. \ref{fig:net}). To mix up $Z'$ using $A$, the decoder feature is simply multiplied by the affinity matrix, i.e. $\tilde{Z} = A Z'$. However, to mix up the encoder feature $X'$, the affinity matrix $A \in [0, 1]^{c \times t \times t}$ needs to be adopted accordingly. First, the matrix $A$ is averaged across the first dimension and then repeated by the first dimension for $b'$ times:
\begin{equation}
\tilde{A}=\mathbbm{1}_{b'}\sum_{i=1}^{c}\exp(A_i)/c
\end{equation}
where $\mathbbm{1}_{b}$ denotes a $b$-dimensional vector of ones and $\tilde{A} \in [0, 1]^{b' \times t \times t}$. The mixing up step is performed by multiplying $\tilde{A}$ and $X'$, i.e. $\tilde{X} = \tilde{A} X'$.
\vspace{-6pt}
\vspace{-6pt}

The \acrshort{am} is closely related to \acrshort{ann} and \acrshort{gnn}. The attention layer (Eq. \ref{eq:att}) can be divided into two encoding parts: the softmax factor responsible for similarity and the output projection $V$. The output part, which is just a linear function, is generalized to any differentiable function in the \acrshort{am}. The dot-product similarity is also replaced with the negative of a distance function. For real-valued spectrograms, in contrast to encoded inputs, the dot-product cannot comply with the baseline without the \acrshort{am}. When the normalized adjacency matrix is available (Eq. \ref{eq:graph}), the mixup step is performed before the nonlinear function, while the \acrshort{am} must be applied after the nonlinear function.

\section{Experiments} \label{sec:experiments}
There are many front-end feature options for the time-frequency representation, each depending on several hyper-parameters. However, \acrshort{sed} is commonly performed using \acrfullpl{lms} because of memory and efficiency concerns. The 64-dimensional \acrshortpl{lms} are extracted by a $2048$ point Fourier transform every $20$ ms with a Hann window size of $40$ ms. In the training, zero padding is applied to the longest sample-length within a batch ($64$ samples), whereas a batch-size of $1$ is used during inference, meaning no padding. The \acrshort{nn} parameters are estimated using AdamW \cite{loshchilov2017decoupled} with a starting learning rate of $1e-4$, and successive learning rate reduction if the cross-validation loss did not improve for three epochs.

\subsection{Dataset}
\begin{figure}[H]
\begin{minipage}[b]{1.0\linewidth}
  \centering
  \centerline{\includegraphics[width=244pt]{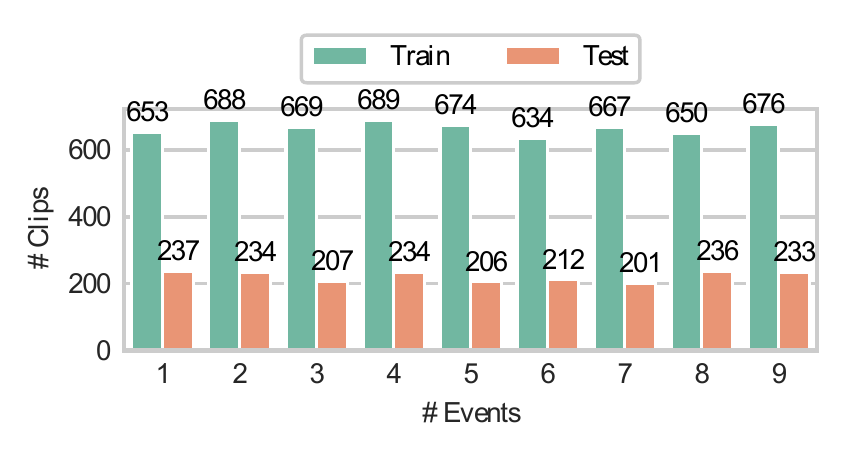}}
\end{minipage}
\vspace*{-25pt}
\caption{The number of events per clip for the train and test set of the URBAN-SED dataset.}
\label{fig:number}
\end{figure}

The URBAN-SED \cite{salamon2017scaper}, a \acrshort{sed} dataset within an urban setting, is a collection of $10$ seconds clips with $10$ event labels. This dataset’s source material is the UrbanSound8k dataset containing $27.8$ hours of data split into $10$-second clip segments. The URBAN-SED dataset encompasses $10,000$ sound scapes generated using the Scaper sound scape synthesis library \cite{salamon2017scaper}, being split into $6000$ training, $2000$ validation and $2000$ evaluation clips. The training set contains mostly $10$-second excerpts, which are weakly labeled, whereas each clip contains between one and nine events. The evaluation dataset is strongly labeled, containing up to nine events per clip. This work only uses the training and evaluation dataset. The number of events per clip for each subset is provided in figure \ref{fig:number}. Note that estimating events with a long duration (e.g., $10$ s) is equal to audio tagging.
\vspace{-6pt}
\vspace{-6pt}



\subsection{Evaluation}
The \acrshort{sed} is evaluated by three different metrics, each represented in terms of F1, precision, and recall. \textbf{Tagging score}: This metric measures the models’ capability to correctly identify the presence of an event within an audio clip. \textbf{Segment score}: This metric is an objective measure of a given model’s sound localization capability, measured by the segment-level (adjustable) overlap between ground truth and prediction. The segment-F1 cuts an audio clip into multiple fixed sized segments \cite{mesaros2016metrics}. The segment-F1 can be seen as a coarse localization metric since precise time-stamps are not required. \textbf{Event score}: This metric measures on- and off-set overlap between prediction and ground-truth, thus is not bound to a time-resolution (like segment-F1). The event-F1 specifically describes a model’s capability to estimate a duration (i.e., predict on- and off-set).

\subsection{Results}
\begin{figure*}[ht]
\begin{minipage}[b]{1.0\linewidth}
  \centering
  \centerline{\includegraphics[width=504pt]{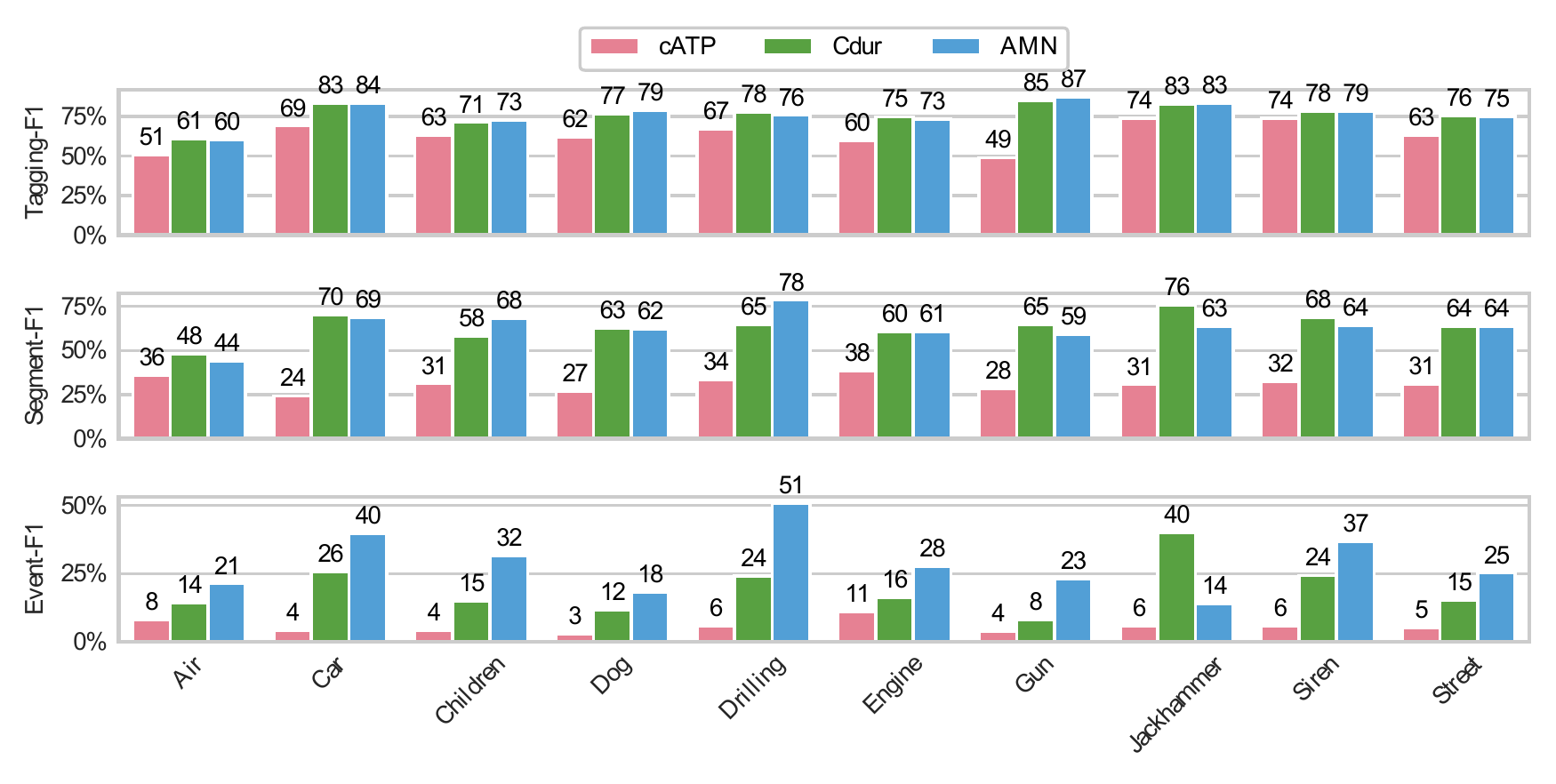}}
\end{minipage}
\vspace*{-25pt}
\caption{\acrshort{sed} per event-F1 score results on the URBAN-SED test dataset. The events are sorted from (left) short to (right) long average duration.}
\label{fig:class}
\end{figure*}

All the results are reported in terms of the average and $95\%$ confidence interval of $10$ experiments. Table \ref{tab:event} compares the event score of \acrshort{amn} with cATP \cite{lin2020specialized} and CDur \cite{dinkel2020duration}, two formerly proposed \acrshortpl{crnn} for \acrshort{wssed}. As showed in that table, our baseline network outperforms all other methods in terms of event-F1.
\begin{table}[ht]
    \centering
    \begin{tabular}{l|c:cc}
        \bf Method & \bf F1 & \bf Precision & \bf Recall \\
        \hline
        cATP & 3.5 $\pm$ 0.7 & 2.0 $\pm$ 0.5 & 25.3 $\pm$ 0.3 \\
        CDur & 20.1 $\pm$ 0.1 & 18.3 $\pm$ 0.2 & 22.5 $\pm$ 0.1 \\
        AMN & \bf 28.3 $\pm$ 0.2 & 34.8 $\pm$ 0.1 & 24.5 $\pm$ 0.2 \\
    \end{tabular}
    \caption{The F1, precision, recall for the \acrshort{sed} in terms of event score.}
    \label{tab:event}
\end{table}
The event-F1 enhancement is a result of improving both precision and recall. However, as shown in Tables \ref{tab:tagging} and \ref{tab:segment}, the improvements of tagging and segment scores only comes from the precision part.
\begin{table}[ht]
    \centering
    \begin{tabular}{l|c:cc}
        \bf Method & \bf F1 & \bf Precision & \bf Recall \\
        \hline
        cATP & 70.0 $\pm$ 2.3 & 73.3 $\pm$ 2 & 67.6 $\pm$ 2.5 \\
        CDur & 76.2 $\pm$ 0.2 & 76.1 $\pm$ 0.3 & 77.1 $\pm$ 0.1 \\
        AMN & \bf 76.8 $\pm$ 0.3 & 79.8 $\pm$ 0.2 & 74.7 $\pm$ 0.3 \\
    \end{tabular}
    \caption{The F1, precision, recall for the audio tagging in terms of tagging score.}
    \label{tab:tagging}
\end{table}
\begin{table}[ht]
    \centering
    \begin{tabular}{l|c:cc}
        \bf Method & \bf F1 & \bf Precision & \bf Recall \\
        \hline
        cATP &  28.7 $\pm$ 0.9 & 16.9 $\pm$ 0.6 & 98 $\pm$ 0.3 \\
        CDur & \bf 63.5 $\pm$ 0.2 & 67.9 $\pm$ 0.3 & 60.8 $\pm$ 0.2 \\
        AMN & 62.9 $\pm$ 0.3 & 75.6 $\pm$ 0.3 & 55.3 $\pm$ 0.4 \\
    \end{tabular}
    \caption{The F1, precision, recall for the \acrshort{sed} in terms of segment score.}
    \label{tab:segment}
\end{table}
Lower tagging-recall means less correct detection and one reason behind this behavior is mixing up a few large values with many small values. For example, consider the maxpool on the frame-level outputs. When the maximum value is slightly higher than threshold, mixing up with rest of the frame reduces the maximum value to somewhere lower than threshold. However, reducing the number of correct predictions does not yield more wrong predictions, which explains higher precision in both tagging and segment scores (Tables \ref{tab:tagging} and \ref{tab:segment}). Moreover, the segment-F1 score of \acrshort{amn} is also slightly higher (63.5\%) than all other previous \acrshort{wssed} methods, where event-F1 is significantly higher (28.3\%). The performance gap between segment-F1 and event-F1 further exemplifies the difficulty in \acrshort{wssed} to obtain fine-scale onset and offset estimates. When adding additional augmentation methods, the performance further improves. Figure \ref{fig:class} shows a comparison between three different models. The class-wise performance is measured for both audio tagging (tagging-F1) and \acrshort{sed} (segment-F1 and event-F1).
Tables \ref{tab:res-enc} and \ref{tab:res-dec} represent the effect of introducing \acrshort{am} on different time resolutions in the encoding and decoding, respectively.
\begin{table}[ht]
    \centering
    \begin{tabular}{l|c:cc}
        \bf Time resolution & \bf F1 & \bf Precision & \bf Recall \\
        \hline
        No AM & 25.7 $\pm $0.4 & 31.6 $\pm$ 0.5 & 22.3 $\pm$ 0.2 \\
        Only $1/2$ & 25.9 $\pm $0.3 & 32.4 $\pm$ 0.5 & 22.3 $\pm$ 0.2 \\
        Only $1/4$ & \bf 27.3 $\pm $0.3 & 34.1 $\pm$ 0.5 & 23.4 $\pm$ 0.2 \\
        $1/2$ and $1/4$ & 26.9 $\pm $0.2 & 33.9 $\pm$ 0.3 & 23 $\pm$ 0.2 \\
    \end{tabular}
    \caption{The effect of applying \acrshort{am} in different layers when used only in the encoding part. \acrshort{sed} is evaluated with F1, precision, recall in terms of event score.}
    \label{tab:res-enc}
\end{table}
\begin{table}[ht]
    \centering
    \begin{tabular}{l|ccc}
        \bf Time resolution & \bf F1 & \bf Precision & \bf Recall \\
        \hline
        No AM & 25.7 $\pm$ 0.4 & 31.6 $\pm$ 0.5 & 22.3 $\pm$ 0.2 \\
        Only $1/2$ & 26.8 $\pm$ 0.3 & 32.6 $\pm$ 0.4 & 23.3 $\pm$ 0.3 \\
        Only $1/4$ & 27.8 $\pm$ 0.4 & 33.9 $\pm$ 0.3 & 24.2 $\pm$ 0.4 \\
        $1/2$ and $1/4$ & \bf 28.3 $\pm$ 0.2 & 34.8 $\pm$ 0.1 & 24.5 $\pm$ 0.2 \\
    \end{tabular}
    \caption{The effect of applying \acrshort{am} in different layers when used only in the decoding part. \acrshort{sed} is evaluated with F1, precision, recall in terms of event score.}
    \label{tab:res-dec}
\end{table}
To explore the importance of the skip connection from the encoder to decoder, Table \ref{tab:res-enc} shows the performance of \acrshort{amn} without \acrshortpl{am} in the decoder. In the encoding, the layer with lowest time resolution has a greatest impact. Without the skip connection, the network with the \acrshort{am} only at the lowest time resolution performs the best. In the presence of the \acrshort{am}-$1/4$, mixing up the $1/2$ resolution degrades the performance which is not the case other way around. However, independent of the time resolution, using \acrshort{am} in any encoding layer increases the event-F1. Also, most of the F1 improvement comes from the precision side. In contrast, the \acrshort{am} in the decoding shows a different behavior compared with the encoding part (Table \ref{tab:res-dec}). The full \acrshort{amn}, including two skip connections to the decoder, achieves the highest event-F1. Similar to the encoding, the lowest resolution ($1/4$) still has a greater impact than other layers and unlike the encoding, the recall part has more contribution to the event-F1. Comparing both AMN-Enc and AMN-Dec with the baseline (without \acrshort{am}), the \acrshort{am} in the encoding improves the event-F1 by $3.8\%$ while it increases the event-F1 by $3.6\%$ when adding to the decoding part.

Table \ref{tab:tau} shows the importance of the hyper-parameter $\tau$. While the \acrshort{amn} indicates an empirical stability to the $\tau$ parameter, the default value $\tau=1$ achieves the best performance.
\begin{table}[ht]
    \centering
    \begin{tabular}{l|ccc}
        $\boldsymbol\tau$ & \bf Tagging-F1 & \bf Segment-F1 & \bf Event-F1 \\
        \hline
        $5.0$ & 76.5 $\pm$ 0.2 & 62.3 $\pm$ 0.3 & 27.5 $\pm$ 0.3 \\
        $1.0$ & \bf 76.8 $\pm$ 0.3 & \bf 62.9 $\pm$ 0.3 & \bf 28.3 $\pm$ 0.2 \\
        $0.5$ & 76.1 $\pm$ 0.2 & 62.2 $\pm$ 0.4 & 27.6 $\pm$ 0.4 \\
        $0.1$ & 76.0 $\pm$ 0.4 & 62.3 $\pm$ 0.6 & 27.5 $\pm$ 0.3 \\
        $0.05$ & 76.2 $\pm$ 0.2 & 62.6 $\pm$ 0.4 & 27.7 $\pm$ 0.5
    \end{tabular}
    \caption{The effect of hyper-parameter $\tau$ on tagging-, segment-, and event-F1.}
    \label{tab:tau}
\end{table}

To explore the role of \acrshort{am} in the optimization step, some gradients are left out of the loss function. It could be seen in Table \ref{tab:grad} that \acrshort{am} heavily depends on the backward gradients. 
\begin{table}[ht]
    \centering
    \begin{tabular}{l|ccc}
        \bf Affinity Grad. & \bf Tagging-F1 & \bf Segment-F1 & \bf Event-F1 \\
        \hline
        No Grad. & 68.4 $\pm$ 0.4 & 38.0 $\pm$ 0.3 & 0.4 $\pm$ 0 \\
        Only Enc. & 76.4 $\pm$ 0.2 & 62.7 $\pm$ 0.5 & 26.7 $\pm$ 0.7 \\
        Only Dec. & 75.8 $\pm$ 0.2 & 61.6 $\pm$ 0.2 & 26.6 $\pm$ 0.2 \\
        Enc. and Dec. & \bf 76.8 $\pm$ 0.3 & \bf 62.9 $\pm$ 0.3 & \bf 28.3 $\pm$ 0.2 \\
    \end{tabular}
    \caption{The effect of \acrshort{am} in the optimization process.}
    \label{tab:grad}
\end{table}
When mixing up features without having their gradients, the affinity matrix is excluded from the loss function. The gradients of affinity in the encoding are more important than decoding. Excluding gradients completely deteriorates the event-F1. The loss function, including the gradients of the affinity matrix in both encoder and decoder, achieves the highest performance.

\section{Conclusion}
\label{sec:conclusion}
In this work, we proposed the \acrfull{am} method for the \acrshort{wssed} problem. The \acrshort{am} was shown to significantly improve the \acrshort{sed} in terms of the F1 score. \acrshort{amn} provides a non-parametric attention mechanism and better time-domain predictions by skip connections from encoder to decoder. Several ablation studies were performed to show that our approach is robust to the hyper-parameter changes, mixup gradients are necessary, and lower resolutions are more important. 
\printbibliography
\end{document}